\documentclass[aps,prl,reprint,groupedaddress]{revtex4-1}

\usepackage{graphicx}% Include figure files
\usepackage{bm}% bold math
\usepackage{verbatim}% bold math
\usepackage[T1]{fontenc}% bold math
\usepackage{textcomp}% bold math
\usepackage{color}
\usepackage[usenames,dvipsnames]{xcolor}

\begin{document}

%\preprint{}

\title{
  Image based cellular contractile force evaluation\\
  with small-world network inspired CNN: SW-UNet
}


\author{Li Honghan}
\author{Daiki Matsunaga}
\email{daiki.matsunaga@me.es.osaka-u.ac.jp}
\author{Tsubasa S. Matsui}
\author{Hiroki Aosaki}
\author{Shinji Deguchi}
\email{deguchi@me.es.osaka-u.ac.jp}
%\thanks{}
\affiliation{%
 Division of Bioengineering, Graduate School of Engineering Science, Osaka University, Japan\\
 1-3 Machikaneyama Toyonaka, Osaka, 5608531 Japan 
}%


\date{\today}

\begin{abstract}
We propose an image-based cellular contractile force evaluation method using a machine learning technique.
We use a special substrate that exhibits wrinkles when cells grab the substrate and contract, and the wrinkles can be used to visualize the force magnitude and direction.
In order to extract wrinkles from the microscope images, we develop a new CNN (convolutional neural network) architecture SW-UNet (small-world U-Net), which is a CNN that reflects the concept of the small-world network.
The SW-UNet shows better performance in wrinkle segmentation task compared to other methods: the error  (Euclidean distance) of SW-UNet is 4.9 times smaller than 2D-FFT (fast Fourier transform) based segmentation approach, and is 2.9 times smaller than U-Net.
As a demonstration, we compare the contractile force of U2OS (human osteosarcoma) cells and show that cells with a mutation in the KRAS oncogne show larger force compared to the wild-type cells.
Our new machine learning based algorithm provides us an efficient, automated and accurate method to evaluate the cell contractile force.
\end{abstract}

%\keywords{}

\maketitle

Cellular contractile force is known to regulate diverse functions, particularly related to cell adhesion, proliferation and migration, thus acting as an essential driver in morphogenesis and pathogenesis \cite{polacheck2016measuring}.
Therefore, measuring cellular contractile force is essential to understand and control the status of living cells.
The most common methods to measure the contractile force are traction force microscopy (TFM) \cite{munevar2001traction} and microneedle assay \cite{tan2003cells,liu2010mechanical}.
In TFM, the displacement field is measured by fluorescent microbeads embedded inside the substrate, and the contractile force is evaluated solving the inverse problem.
In microneedle assay, the contractile force is evaluated from the deflections of the microneedles on which cells are plated.

\begin{figure*}
    \includegraphics[width=1.4\columnwidth]{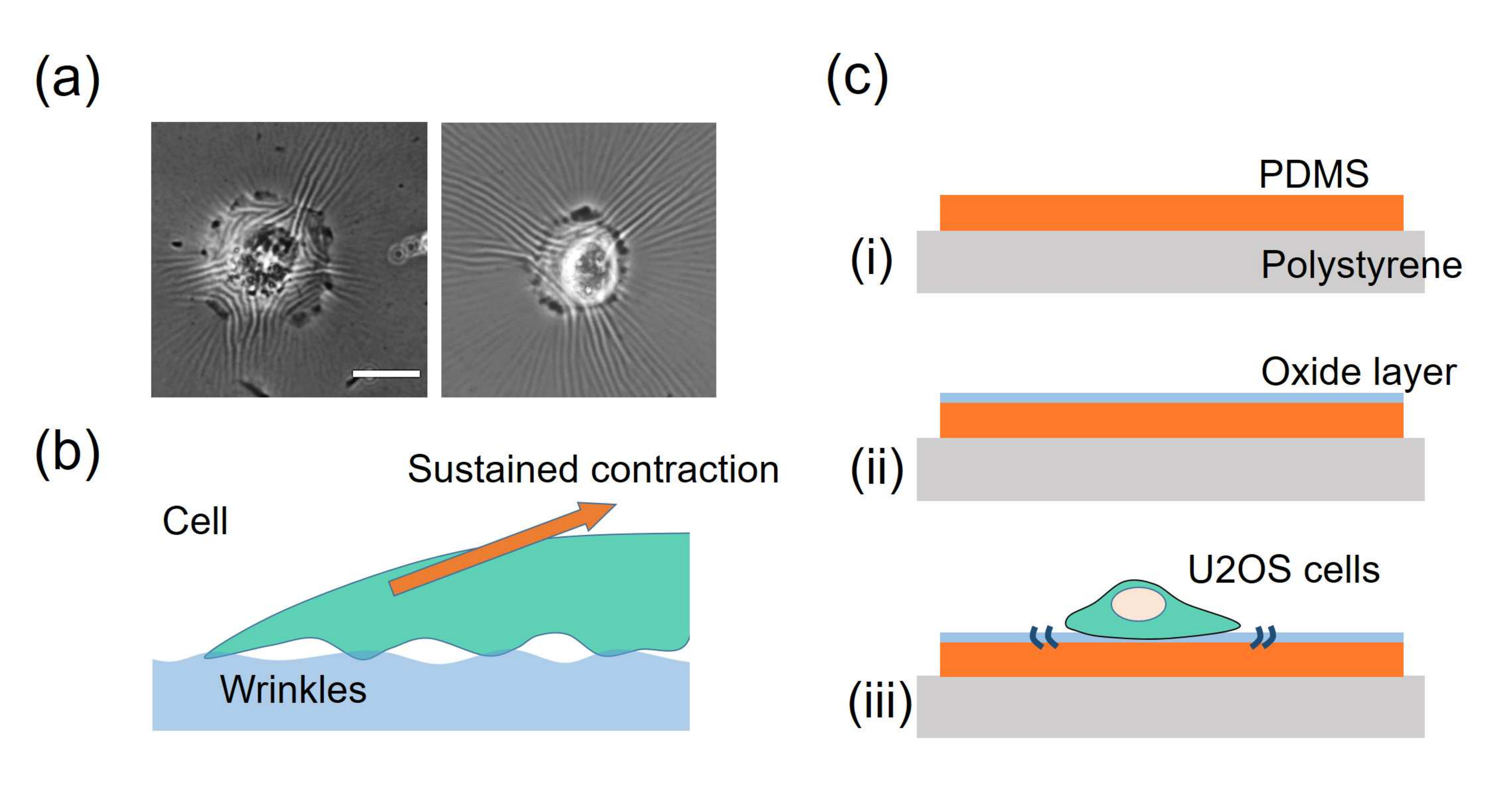}
    \caption{{\bf Wrinkle generation by the cell contractile force and fabrication method of the substrate:}
    (a) Microscope images of wrinkles that are generated by the U2OS cell contractile force. The scale bar in the figure has a length of 20 micrometers.
    (b) Schematic side view of the cell. The contractile forces are generated by cellular endogenous activity, and the force gives rise to the wrinkle generation.
    (c) Schematics of our experiment procedures. 
    (i) The PDMS gel layer is coated on the polystyrene layer.
    (ii) The oxygen plasma is applied to the PDMS gel layer to oxide the surface layer. 
    (iii) The U2OS cells are cultured on the substrate.
    \label{fig1}}
    \centering
\end{figure*}

Another method used to evaluate the contractile force is the wrinkle based measurements \cite{burton1997traction,balaban2001force,yokoyama2017new,ichikawa2017vinexin,fukuda2017cellular}.
In a special substrate that has a stiff top layer by heating \cite{harris1980silicone,burton1997traction} or plasma irradiation \cite{ichikawa2017vinexin,fukuda2017cellular}, cells generate wrinkles when they grab the substrate and contract as shown in Fig.~\ref{fig1}(a)(b), and the wrinkles can be used to visualize the force magnitude and direction.
The wrinkle length can be used to estimate the force magnitude since the wrinkle length has a positive correlation with the force strength \cite{balaban2001force,fukuda2017cellular}.
In previous studies, researchers tried to extract the wrinkles and measure its length manually \cite{balaban2001force}, or by 2D-FFT (fast Fourier transform) based image processing \cite{ichikawa2017vinexin,fukuda2017cellular}.
Although the wrinkle based measurement provides a convenient and efficient way to evaluate the contractile force, it was difficult to extract the wrinkle from the microscope images both accurately and automatically. 
In this work, we proposed a CNN (convolutional neural network) based method to automate segmentation of wrinkles from the microscope images.

In recent years, U-Net \cite{ronneberger2015u,falk2019u} is widely used in the segmentation task for biomedical images including those of cells \cite{van2016deep,fabijanska2018segmentation,niioka2018classification}. 
In this paper, we propose a new CNN called SW-UNet (small-world U-Net), which is a modified U-Net that reflects the concept of the small-world network \cite{watts1998collective,humphries2008network,neal2017small}.
The small-world network is a network that has more connection to its neighbouring nodes while they have less connection to non-neighbouring nodes, and this attribute can be quantified by SWI (small-world index) \cite{neal2017small}.
The original CNN algorithm was initially inspired by the neural structure of the striate-cortex from macaques and cats \cite{hubel1968receptive}.
Since the attribute of the small-world network also exists in the neural structure of animal cortex \cite{bullmore2009complex,rubinov2010complex,sanz2010loss}, we hypothesize that integrating this attribute and building SW-UNet will improve the performance of CNN.
In this work, we built our CNN based on the structure of U-Net and optimized the connection to reflect the concept of the small-world network.
Although there are several recent studies \cite{xie2019exploring,javaheripi2019swnet} worked on image classifications or recognitions based on the small-world inspired CNN, our work is one of the first attempts to work on the image segmentation for a practical application.
Our work is also important because we provide comprehensive knowledge how the network structure affects the segmentation performance.

This paper consists of the following four parts.
Firstly, we prepare training datasets for SW-UNet using image processing techniques.
Secondly, we construct the SW-UNet architecture by importing the attribute of a small-world network into U-Net.
Thirdly, we compare the accuracy of wrinkle extraction with other methods.
Finally, we apply this novel technology to demonstrate that the contractile force in U2OS (human osteosarcoma) cells is elevated upon a mutation in the KRAS oncogene.

\section*{Experimental materials}
\paragraph*{Cell substrate}
Based on our previous studies \cite{yokoyama2017new, fukuda2017cellular}, we prepare the substrate that can generate wrinkles reversibly upon application of cellular forces following steps as in Fig.~\ref{fig1}(c).
Firstly, parts A and B of CY 52-276 (Dow Corning Toray) are mixed at a weight ratio of 1.25:1 to form a PDMS (polydimethylsiloxane) gel layer that is coated on a circular cover glass.
Secondly, the cover glass is placed in a 60\textdegree{}C oven for 20 hours to cure the PDMS gel.
Thirdly, oxygen plasma (SEDE-GE, Meiwafosis) is applied uniformly along the surface of the PDMS layer to create an oxide layer that works as the substrate for cell culture.
Finally, the substrate is coated with 10 μg/mL collagen type I solution for 3 hours. 

\paragraph*{Cells}
U2OS cells (HTB-96; ATCC) were maintained in DMEM (043-30085; Wako) supplemented with 10\% FBS (SAFC Bioscience), 100 U/mL penicillin, and 100 µg/ mL streptomycin (168-23191; Wako). Cells were maintained in a humidified 5\% CO$_2$ incubator at 37\textdegree{}C. 

\paragraph*{Plasmids}
The human KRAS wild-type cDNA (Addgene plasmid \#83166, a gift from Dominic Esposito) and KRAS G12V cDNA (Addgene plasmid \#83169, a gift from Dominic Esposito) were amplified using KOD-plus-Neo DNA polymerase kit (KOD-401; Toyobo).
The expression plasmids encoding mClover2-tagged KRAS wild-type and mRuby2-tagged KRAS G12V were constructed by inserting the PCR-amplified cDNAs into the mClover2-C1 vector (Addgene plasmid \#54577, a gift from Michael Davidson) and the mRuby2-C1 vector (Addgene plasmid \#54768, a gift from Michael Davidson).
Before seeding two populations of KRAS expressing cells onto the gel substrate, cells were transiently transfected with either mClover2-KRAS wild-type or mRuby2-KRAS G12V using ScreenFect A (299-73203; Wako).

\section*{Methods}
\subsection*{Overview}
\begin{figure*}
        \includegraphics[width=1.6\columnwidth]{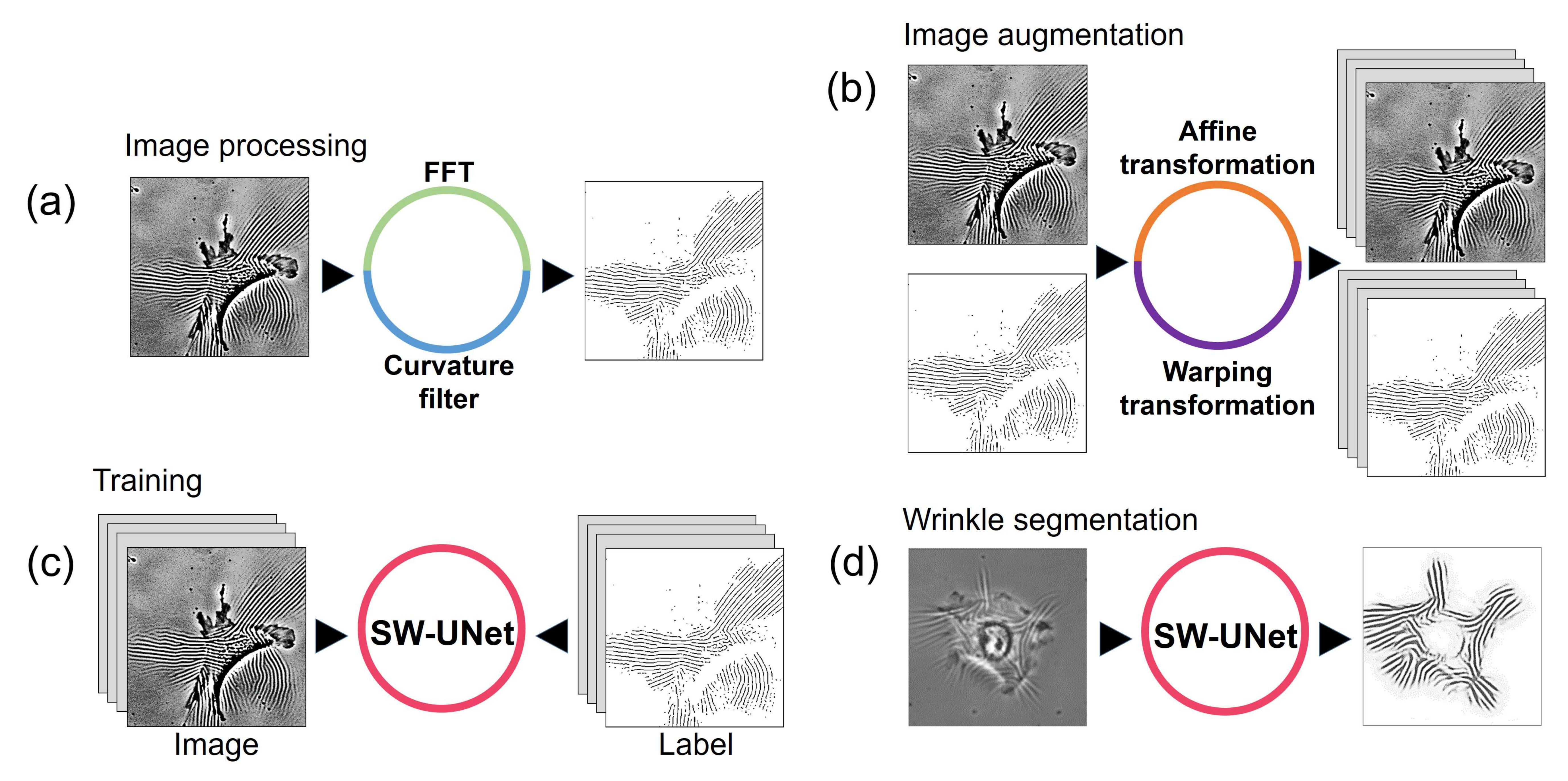}
        \caption{{\bf Overview of our approach.}
        {
        (a) Preparation of training dataset. The wrinkles are extracted by image processing techniques, 2D-FFT (bandpass filtering) and curvature filter.
        (b) Image augmentation methods, affine and warping transformation, are used to increase the number of the training dataset.
        (c) Training SW-UNet from two images: the original microscope images and extracted wrinkle images.
        (d) Utilize SW-UNet to extract wrinkles.
        } \label{fig2}}
\end{figure*}

We overview our CNN-based wrinkle detection system in Fig~\ref{fig2}.
The full process consists of these three steps: (a)-(b) preparing the training dataset, (c) training and (d) wrinkle segmentation.
Firstly, we utilize 2D-FFT method \cite{ichikawa2017vinexin} and curvature filter \cite{Gong2017} to extract rough wrinkle images for the CNN training, as shown in Fig.~\ref{fig2}(a).
Note images of cells and wrinkles are captured on an inverted phase-contrast microscope (IX73; Olympus) using a camera (ORCA-R2; Hamamatsu) with a 20$\times$ objective lens.
A large number of cells cultured on the same substrate were imaged almost simultaneously using an XY motorized stage (Sigma Koki).
In this step, the wrinkles are detected purely by the image processing techniques, and image augmentation is used to increase the number of training data.
Secondly, we train SW-UNet using images that we prepared in the first step: raw cell image (input) and wrinkle image (label) shown in Fig.~\ref{fig2}(c).
Finally, we utilize this SW-UNet to obtain the wrinkles from test images as in Fig.~\ref{fig2}(d). 
In the following subsections, we explain each step in detail.

\subsection*{Training dataset preparation}
\begin{figure*}
        \includegraphics[width=1.4\columnwidth]{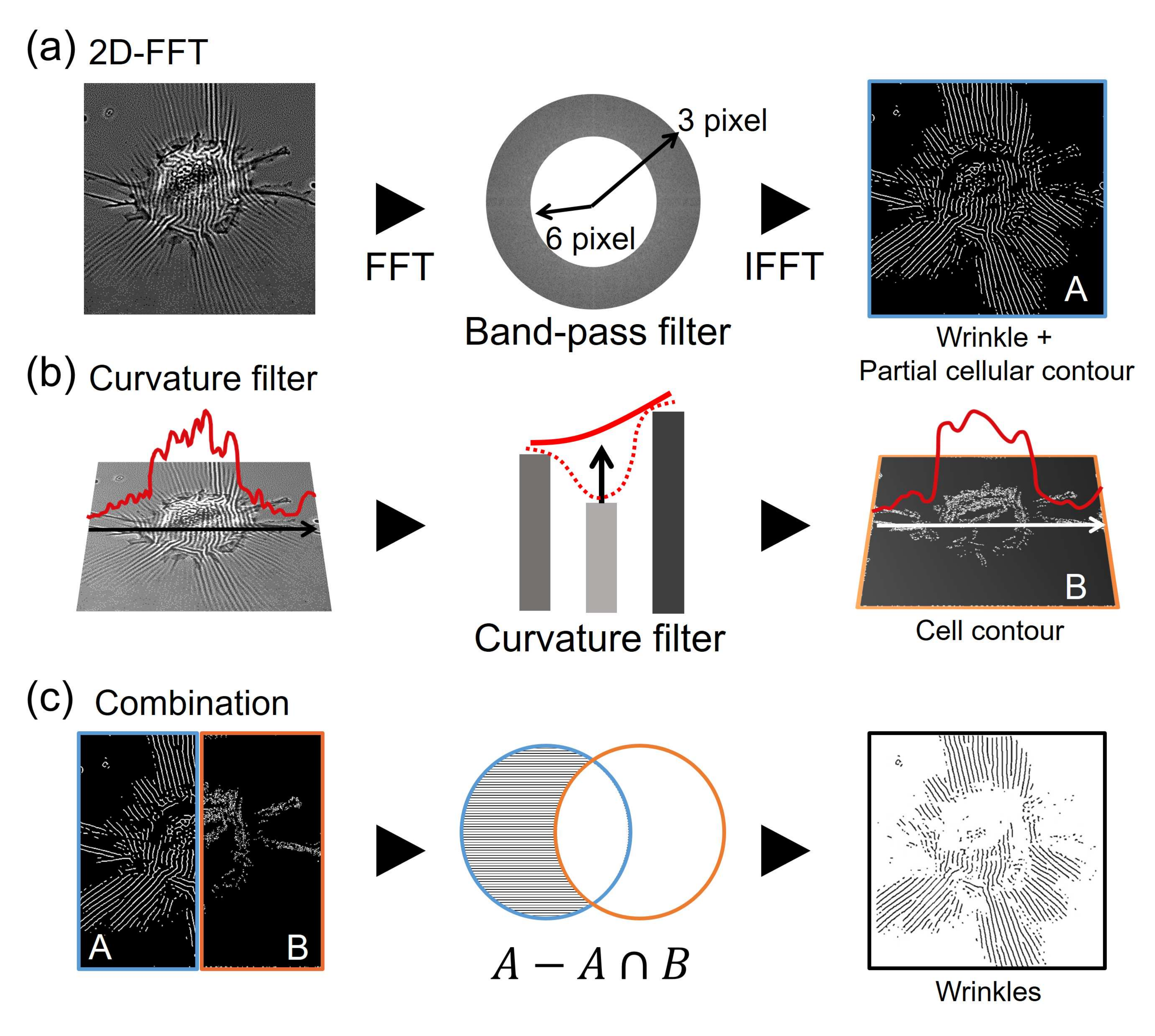}
        \caption{{\bf Preparation of training data: wrinkle extraction with image processing techniques.}
        (a) Rough extraction of wrinkles by a combination of three operations: 2D-FFT, bandpass filtering and IFFT.
        Since the wrinkles have their characteristic wavelength (3-6 pixels), they can be extracted (bandpass filtering) and restored (IFFT) with these three steps. 
        (b) Extracting cell contours from the original images utilizing the curvature filter. Smoothing out the wrinkles, which has a smaller wavelength (i.e. high curvature), the cell contour is extracted. 
        (c) Constructing clear wrinkle image combining two resultant images A and B. 
        \label{fig3}}
  \centering
\end{figure*}

\paragraph{{2D-FFT and bandpass filter}}
The wrinkle patterns are firstly extracted by combinations of successive three operations: 2D-FFT, bandpass filtering and inverse FFT (IFFT) techniques \cite{fukuda2017cellular, ichikawa2017vinexin}. 
Note this approach has been already established and utilized in our previous studies \cite{fukuda2017cellular, ichikawa2017vinexin}, and please refer to these papers for details.
Since the wrinkles have a characteristic wavelength (3-6 pixels), the pattern can be extracted applying a bandpass filter to the image after the 2D-FFT operation as shown in Fig.~\ref{fig3}(a). 
Restoring the image with IFFT, the wrinkles can be extracted as the figure, but the image also contains cell contours.

\paragraph{Curvature filter}
Curvature filter is originally designed to achieve efficient smoothing and denoising operations \cite{Gong2017}. 
Considering the image intensities as a heightfield, the surface mean curvature can be obtained at each pixel.
The filter can be used to smooth out only wrinkles because pixels that have higher curvature decay faster in this filter.
Figure~\ref{fig3}(b) shows images before and after the curvature filter, and it is clearly shown that the wrinkles smoothed out, and only cell contours remained.
Note we utilized the filter repeatedly 200-1000 times until only wrinkles disappear. 

Computing conjunction ($A \cap B$) of two resultant images, $A$ (right end of Fig.~\ref{fig3}(a)) and $B$ (right end of (b)), the cell contours that appear in image $A$ can be extracted.
Finally substituting the cell contours ($A \cap B$) from image $A$ as shown in Fig.~\ref{fig3}(c), images with only wrinkles are obtained. 

\paragraph*{Image augmentation}
We prepared 126 original cell images for the training.
Many previous researches that handle biomedical images \cite{ronneberger2015u,wu2015deep} used image augmentation techniques to increase the number of training images.
In this study, we also expand the quantity of our cell images from 126 to 1404 by the geometric affine transformations \cite{wang2017effectiveness,gao2016hep} and warping transformations.

\subsection*{CNN architecture}
Although the traditional image processing techniques are effective as shown in the previous section, the method fails to reproduce the wrinkle pattern in some cases (as also shown later in Fig.~\ref{fig6}(a)).
This image processing approach is not applicable in following three situations: (i) when the wrinkles are entirely underneath and overlapped with the cell, (ii) when the wrinkles have fewer features of wave-like patterns and (iii) when there are intense noises in the images.
In this work, we utilize CNN to overcome the situation and to extract clear wrinkle images.

\begin{figure*}
        \includegraphics[width=1.6\columnwidth]{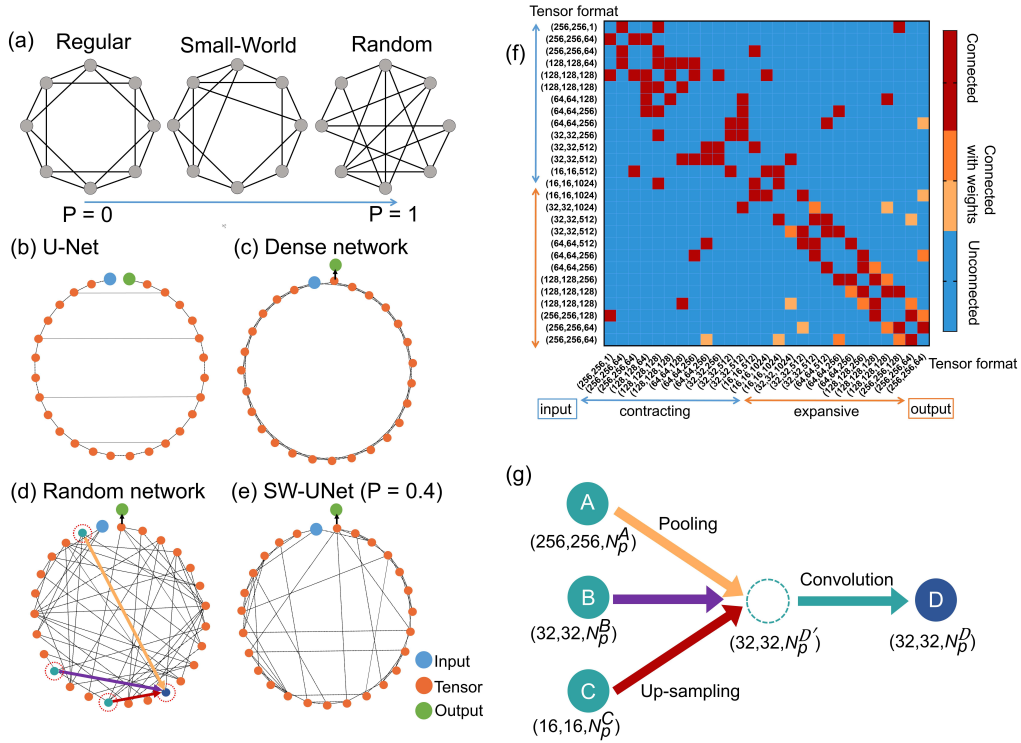}
        \caption{{\bf Overview of SW-UNet architecture.}
         (a) Network topology difference based on the random re-connection probability $P$.
         (b)-(e) Network topology of several CNN structures. Each node corresponds to tensor format, while black lines correspond to the tensor conversions.
         (f) A table showing the node connection status for (e) SW-UNet. Labels on the horizontal and vertical axis are both tensor formats, and the colors inside the table represent the connection status: red shows connected nodes, blue shows unconnected nodes and orange shows connected nodes but with the recursively reduced number of input images.
         (g) A schematic showing a tensor conversion with three input nodes $A-C$ and a single output node $D$.
         \label{fig4}}
\centering
\end{figure*}

In recent researches, U-Net \cite{ronneberger2015u} has been widely used for segmentation of biological and medical images \cite{han2018framing,kayalibay2017cnn,dong2017automatic}.
Figure \ref{fig4}(b) shows network topology of U-Net, and each node corresponds to the tensor format ($N_x, N_y, N_p$); $N_x$ and $N_y$ represent the image size in pixel units both $x$- and $y$-direction respectively, while $N_p$ is the number of images.
Starting from a single input image ($N_x, N_y$, $N_p = 1$), which is shown with a blue node in Fig.~\ref{fig4}(b), the input image goes through the network counter clockwise.
Lines between the nodes are the tensor conversions, such as the pooling and convolution operations.
The image would finally come back to a single output image ($N_x, N_y$, 1) at the green node, and the network is designed to extract the desired segmented image at this final tensor.

The U-Net mainly consists of two paths, contracting path (left side of Fig.~\ref{fig4}(b)) and expansive path (right side).
The contracting path is responsible for extracting the feature from the images, while the expansive path is designed to reconstruct the desired object from the image features.
The contracting path shrinks the image size using the alternate operations of convolution and pooling in the order of (pooling, convolution, convolution).
As the result of these procedures, $N_x$ and $N_y$ decrease while $N_p$ increases.
On the other hand, the expansive path increases the image size $N_x$ and $N_y$ while decreasing $N_p$ using alternate operations of (upsampling, convolution, convolution). 
The image sizes $N_x$ and $N_y$ reach to a minimal after the contracting path, and come back to the original size after the expansive path. 
There are special bypass connections in U-Net called ``copy and crop" path \cite{ronneberger2015u}, which goes horizontally from the contracting to the expansive path in Fig.~\ref{fig4}(b), and the path is responsible for avoiding the loss of effective information during the operation of pooling.

\subsubsection*{Algorithm building SW-UNet}
We now introduce the concept of the small-world and modify the CNN topology.
The topology of the small-world network is characterized and controlled by three parameters: $N$, $K$ and $P$ \cite{watts1998collective,kochen1989small}: $N$ is the number of nodes in the network, $K$ is the average number of the connection branches between the neighbouring nodes, and $P$ is the random reconnection probability. 
The total number of branches is $KN/2$, and selected $\sim KNP/2$ branches are randomly re-connected to other nodes in the network.
Figure~\ref{fig4}(a) shows the schematic of the small-world network topology under fixed $N = 8$ and $K = 4$, but different $P$ parameter.
Each node has connections only to its closest neighbouring $K$ nodes for $P = 0$, and the network topology becomes disordered with the increase of $P$.
We built our SW-UNet architecture through the following procedures.

\paragraph*{Network topology generation} 
In the first step, we build the DenseNet \cite{huang2017densely,jegou2017one,li2018imaging} with $N = 27$, $K = 4$ as shown in Fig.~\ref{fig4}(c).
Each node corresponds to a tensor format ($N_x, N_y, N_p$), and the input image would go through the network counter-clockwise as U-Net.
Following the tensor conversions of U-Net, SW-UNet also consists of the contracting path with successive operations of (pooling, convolution, convolution) and the expansive path with (upsampling, convolution, convolution).
Figure~\ref{fig4}(f) shows the list of tensor formats that we use in SW-UNet.

In the second step, we reconnect randomly selected $\sim NKP/2$ connections for $P \neq 0$ as shown in Fig.~\ref{fig4}(d)-(e). 
The network is DenseNet for $P = 0$, while the network is totally random for $P = 1$ as shown in Fig.~\ref{fig4}(e). 
The image flow direction is always from the upstream to the downstream node.

\paragraph*{Node connection}
The format conversions are necessary to connect nodes that have different tensor formats, and Fig.~\ref{fig4}(g) is a schematic of our connection algorithm.
The extracted connections are from Fig.~\ref{fig4}(d), and it shows a situation that three input nodes $A-C$ are connected to a single output node $D$. 
We first use the pooling and up-sampling operations to match the image size of destination node $D$, $N_x^D = N_y^D = 32$.
For example, the pooling operation is utilized to contract large images as node $A$ ($N_x^A = N_y^A = 256$), while up-sampling operation is utilized to expand smaller images as node $C$ ($N_x^C = N_y^C = 16$).
Summing up all resultant images from node A-C, the number of total images is now $N_p^{D'}=N_p^A+N_p^B+N_p^C$
but the value $N_p^{D'}$ would not necessary match the destination node image number $N_p^{D'} \neq N_p^{D}$.
Therefore, the convolution operation is utilized to convert the image number from $N_p^{D'}$ to $N_p^D$.
Note when one of the input image number ($N_p^A$, $N_p^B$ and $N_p^C$) exceeds the destination image number $N_p^D$, we halve the input image number recursively until they become smaller than $N_p^D$.
Figure~\ref{fig4}(f) shows the connection status for the network $P = 0.4$ (Fig.~\ref{fig4}(e)): red shows connected nodes, blue shows unconnected nodes, and orange shows connected nodes but with the recursively reduced number of input images.

\subsection*{Training parameter}
The number of the training dataset is 1404 (126 original images), and Adam optimizer \cite{kingma2014adam} with a learning rate of 0.0001 is utilized for training the CNN network.
We used Nvidia Titan Black (2 GPUs) to accelerate the training process.

In previous studies, researchers prepared original images in an order of $\sim 1000$ \cite{roth2014new, depeursinge2012building,shin2016deep,zhang2016automatic,sampaio2011detection} as the training dataset to avoid the overfitting.
Since we have 126 original images for the training dataset, we need to restrict our training epochs \cite{loughrey2004overfitting}.
Therefore, we set the training steps in one epoch as 300 and set the total epochs as 10. 

\subsection*{Wrinkle evaluation}
\begin{figure*}
  \includegraphics[width=1.6\columnwidth]{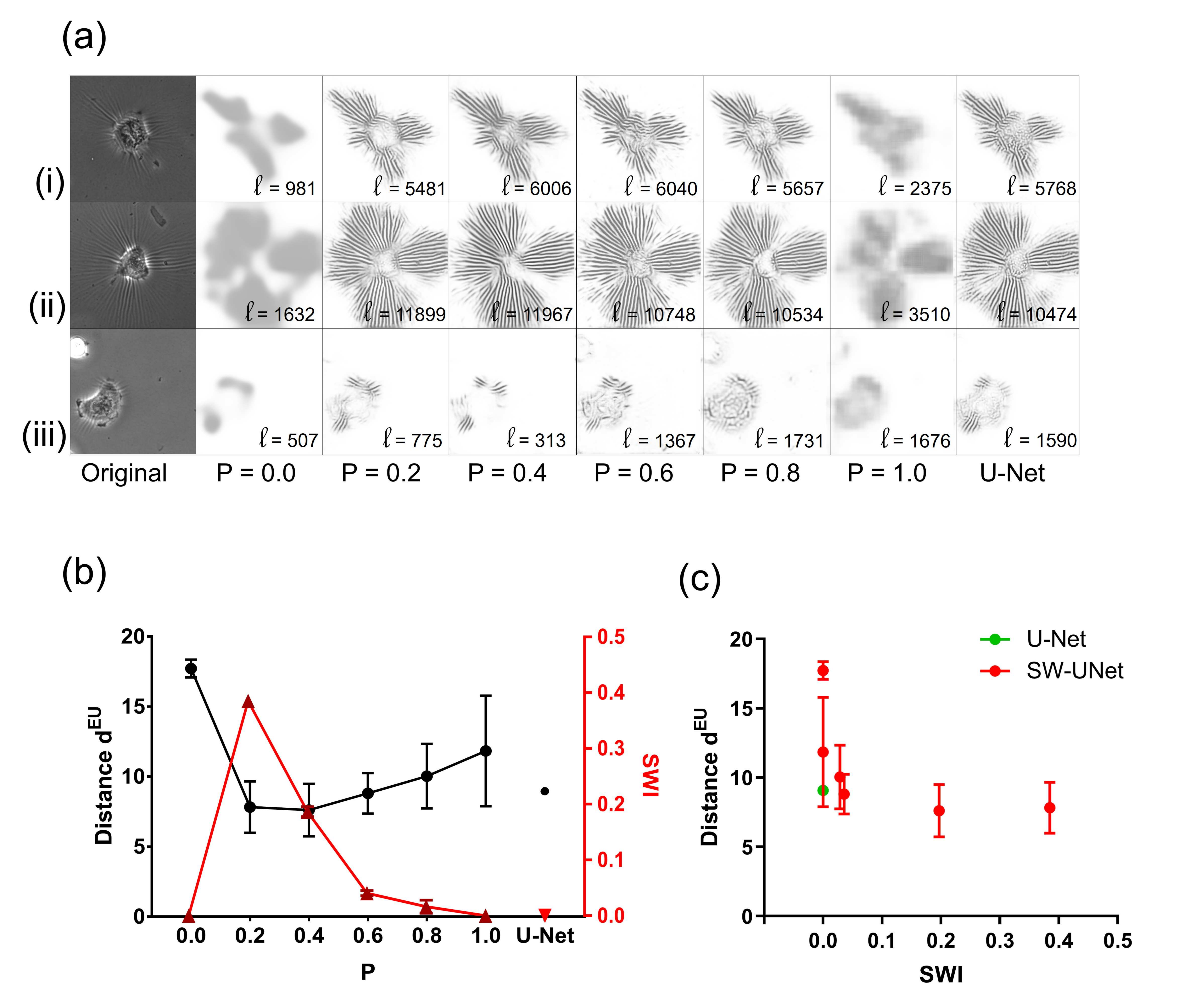}
  \caption{{\bf \textbf{Wrinkle segmentation using SW-UNet with different $P$-values.}}
  (a) Wrinkle images that are produced by SW-UNet and U-Net, and $\ell$ is the wrinkle perimeter length.
    (b) Black plots show the distance $d^{\rm EU}$ while red plots show SWI (\ref{eq:swi}) of the network, as a function of network reconnection parameter $P$.
  (c) The distance $d^{\rm EU}$ as a function of $SWI$, and the figure indicates that our network SW-UNet achieves better performance for larger $SWI$. \label{fig5}}
\end{figure*}

After training CNNs, we evaluate its accuracy with $N_{\rm test} = 58$ test images by comparing with the ground-truth data.
The ground-truth data are produced by three different researchers that were asked to trace the wrinkle lines manually.
Although the cross-entropy is the standard method to compare images \cite{shore1980axiomatic,yi2004automated,ronneberger2015u,shin2016deep},
we did not use this method because it was not a proper criterion to compare the performance of different networks. 
Interestingly, the accuracy (range: 0.9642-0.9759) and loss (range: 0.798-0.808) in the training process converge almost to a same value for all networks, though there is a significant difference in the extracted wrinkles (as shown in Fig.~\ref{fig5}(a)).

Instead, we utilize perimeter length of the wrinkles $\ell$ as the comparison criteria. 
In order to obtain the perimeter, we extract the wrinkle edge with the Prewitt operator at a threshold of 0.01 and count up the number of edge pixels to obtain $\ell$. 
We introduce two different distances, Euclidean $d^{\rm EU}$ and cosine distance $d^{\rm COS}$, to quantify the difference between the wrinkle perimeter obtained by CNN $\ell^{\rm CNN}$ and the ground truth $\ell^{\rm GT}$.
Each distance is defined as
\begin{eqnarray}
	d^{\rm EU}  &=& \sqrt{\sum_{i=1}^{N_{\rm test}} (\ell^{\rm CNN}_i - \ell^{\rm GT}_i)^2}, \label{eq:ed} \\
	d^{\rm COS} &=& 1 - {\frac{{\displaystyle \sum_{i=1}^{N_{\rm test}}} (\ell^{\rm CNN}_i \cdot \ell^{\rm GT}_i)}{\displaystyle \sqrt{{\sum_{i=1}^{N_{\rm test}}} (\ell^{\rm CNN}_i)^2}\sqrt{{\sum_{i=1}^{N_{\rm test}}} (\ell^{\rm GT}_i)^2}}}.
	\label{eq:cod} 
\end{eqnarray}

\section*{Results}
\subsection*{Effect of $P$-value in SW-UNet}
We first evaluate the segmentation performance using different network topology, SW-Net ($P = 0$ to 1) and U-Net, in Fig.~\ref{fig5}(a).
Although most of the networks succeeded in extracting the wrinkles to some extent, $P = 0$ (DenseNet) and $P = 1$ (SW-UNet) failed, and they only showed vague regions of wrinkles.
Comparing the wrinkle perimeter length $\ell$ for different SW-UNets, images (i) and (ii) shows maximum length at intermediate $P = 0.4-0.6$, while image (iii) shows larger $\ell$ for larger $P$-values. 
For images (i) and (ii), the wrinkles are well extracted in $P=0.4-0.6$ but become less prominent with $P$ increase.
As a result, SW-UNets with large $P$-value would underestimate the wrinkle length.
In the case of image (iii), the network with $P = 0.6-0.8$ overestimates the wrinkle length because the network failed to distinguish the cell contours and wrinkles. 
Figure~\ref{fig5}(b) shows the distance $d^{\rm EU}$ from the manually tracked ground truth, and the result shows that the segmentation performance is best at $P = 0.2-0.4$. 
The distance of U-Net was almost the same as SW-UNet with $P = 0.6$.

We now introduce SWI (small-world index) \cite{neal2017small} to characterize the network topology, which is defined as
\begin{equation}
    SWI = 1 - (\frac{L-L_l}{L_r-L_l}-\frac{C-C_r}{C_l-C_r}) \label{eq:swi}
\end{equation}
where $L$ is the average path length and $C$ is the clustering coefficient defined as
\begin{eqnarray}
    L &=& \frac{1}{N(N - 1)} \sum_{i}^N \sum_{j \neq i}^N D_{ij}, \label{eq:l} \\
    C &=& \frac{1}{N} \sum_{i}^N \frac{\displaystyle \sum_j^N \sum_k^N a_{ij} a_{ih} a_{jh}}{\displaystyle (\sum_j^N a_{ij})\cdot(\sum_j^N a_{ij}-1)}. \label{eq:c}
\end{eqnarray}
$D = 1$ is the distance between two nodes, $N = 27$ is the number of nodes in the network and {$a$ is the connection status between two nodes: $a_{ij} = 1$ when nodes $i$ and $j$ are connected while $a_{ij} = 0$ if the nodes are not connected}.
Subscripts $l$ and $r$ describes that the value is from the regular or random network respectively: $C_{l}$ and $C_{r}$ are the clustering coefficients for regular and random networks, while $L_{l}$ and $L_{r}$ are the average path lengths in regular and random networks.

Figure~\ref{fig5}(b) shows that SWI reaches maximum at $P = 0.2$ and gradually decrease with $P$ increase.
Plotting distance $d^{\rm EU}$ as a function of SWI as shown in Fig.~\ref{fig5}(c), the result infers that the network with larger SWI has better segmentation performance. 
Note we evaluated the distance and SWI with three randomly generated network for each $P$ value.
In recent years, there was a report on the macaques and cats cortex topology \cite{sporns2004small}, and the small-world index was estimated as $SWI \approx 0.4$ from their results.
The network topology in the brain might be optimized in the process of evolution.
Although we cannot draw a definite conclusion here because of the small number of sample data, there is a possibility that the network SWI is one criterion to judge the performance when designing a new CNN.

From next sections, we will fix the value to $P = 0.4$ for SW-UNet.

\begin{figure*}
    \includegraphics[width=1.6\columnwidth]{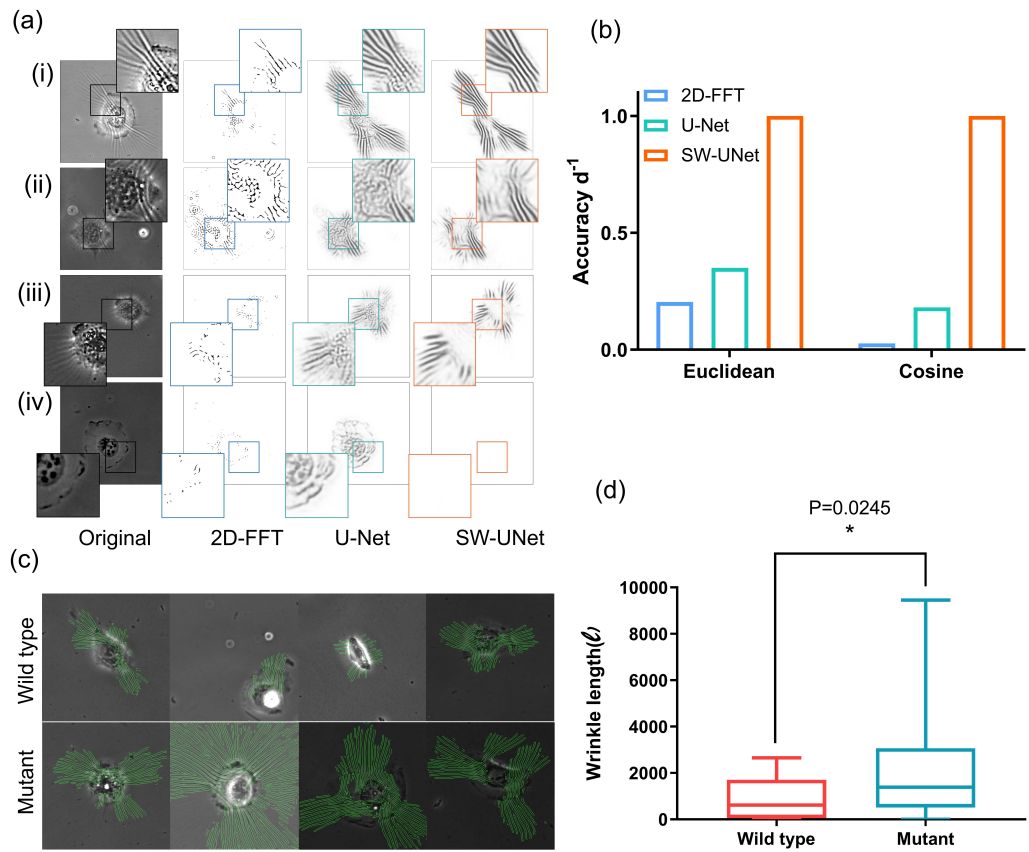}
    \caption{
    \textbf{
    Wrinkle segmentation accuracy of SW-UNet and its application.
    }
    (a) Comparison of extracted wrinkles by different methods.
    (b) Accuracy of wrinkle segmentation quantified by the distances, Euclidean and cosine distances, from the ground truth data. 
    SW-UNet has the smallest error compared to 2D-FFT based segmentation and U-Net.
    (c) The wrinkles (green lines) extracted from the microscope images by SW-UNet for U2OS cells with mutant KRAS gene (first row), and wild-type U2OS cells (second row).
    (d) Wrinkle lengths of the two cell types. The mutant cell has longer wrinkle compared to the wild- type, and there is a significant difference (student's $t$-test) in two groups.
    \label{fig6}
    }
\centering
\end{figure*}

\subsection*{Comparison of different segmentation methods}
Figure~\ref{fig6}(a) compares extracted wrinkles with different approaches: 2D-FFT based method (image processing based segmentation), U-Net and our SW-UNet.
The 2D-FFT based method has the worst segmentation performance, and extracted wrinkles are dotted-line-like patterns rather than continuous lines.
This is because the 2D-FFT based method can only detect the patterns that have periodic wave patterns, and it has a limitation detecting complex-shaped wrinkles as images (ii) or (iii). 
The third row of Fig.~\ref{fig6}(a) shows the images generated by U-Net.
Although the wrinkles are extracted clearer compared to the 2D-FFT based approach, U-Net failed to distinguish the cell contours and wrinkles in some circumstances.
For example, U-Net treated the cell organelles as the wrinkles in images (ii) and (iii) and accordingly overestimating the length of wrinkles.
In the case of image (iv), U-Net detected wrinkles at the cell perimeter even though there are no apparent wrinkles in the microscope image.
On the other hand, SW-UNet succeeded in distinguishing the wrinkles from the cell contour, and the wrinkle length can be evaluated precisely.

We now introduce the Euclidean distance (\ref{eq:ed}) and cosine distance (\ref{eq:cod}) to quantify the segmentation accuracy.
Figure~\ref{fig6}(b) shows the accuracy, which is the inverse of the distance $1/d$, obtained by comparing with manually traced wrinkle lines.
Note the accuracy $1/d$ is normalized by the score of SW-UNet in the figure.
The figure shows that SW-UNet has far better performance compared to other two approaches, and the accuracy based on Euclidean distance $1/d^{\rm EU}$ was 4.9 times accurate compared to the 2D-FFT based approach, and 2.9 times accurate compared to U-Net. 
In the case of the accuracy based on cosine distance $1/d^{\rm COS}$, it was 36.8 times accurate compared to 2D-FFT based approach, and 5.5 times accurate compared to U-Net.
In summary, our SW-UNet is the most effective method for this application. 

\subsection*{Demonstration: Effect of KRAS mutation}
To demonstrate that our SW-UNet is applicable to evaluate the cellular contractile force, we finally evaluate the force with and without a KRAS mutation and compare them.
Mutations in the KRAS oncogene are highly correlated with various types of cancer development \cite{tsuchida2016kirsten}, including metastatic colorectal cancer \cite{amado2008wild}, pancreatic cancer \cite{son2013glutamine} and non-small cell lung cancer \cite{riely2009kras}.
G12V, which is a point mutation with a replacement from glycine to valine at amino acid 12, is one of the most common oncogenic KRAS mutations and has been reported to result in enhanced myosin phosphorylation \cite{hogan2009characterization}.

Utilizing our new SW-UNet method, we extracted the wrinkles from the microscope images, as shown in Fig.~\ref{fig6}(c), and the mutant group shows more wrinkles than the wild-type group.
In supplemental meterial, we also show movies of moving cells with extracted wrinkles (Movie 1 and 2). 
Figure~\ref{fig6}(d) compares the wrinkle length $\ell$, and the average length of mutant cells ($\ell = 2144$) is larger than that of the wild-type ($\ell = 901$).
Student's t-test shows that the $p$-value between these two groups is 0.0245, and thus indicating that the mutant group and wild-type group are significantly different. 
The previous study \cite{hogan2009characterization}, which reported enhanced myosin phosphorylation upon G12V mutation, indirectly suggests an increased force generation during cancer development.
In accordance with this study, our present result demonstrates that the mutated cells indeed exhibit greater forces.

Given that comprehensive analyses are often crucial in the field of cell biology to evaluate, e.g., how mutations in specific oncogenes or administration of specific drugs result in changes in cellular physical forces, our system with SW-UNet of high-throughput capability is potentially useful to more thoroughly evaluate potential changes in the cellular contractile force upon different types of molecular perturbations. 

\section*{Conclusion}
In this paper, we proposed an image-based cellular contractile force evaluation method using a machine learning technique.
We developed a new CNN architecture SW-UNet for the image segmentation task, and the network reflects the concept of the small-world network.
The network topology is controlled by three parameters: number of nodes $N$, number of connection branches from a single node to other $K$ and re-connection probability $P$.
Our network reaches to the maximum segmentation performance at $P = 0.2-0.4$, and the result infers that the networks with larger SWI might have better performance in the segmentation. 
Using our SW-UNet, we can extract the wrinkles clearer than other methods.
The error (Euclidean distance) of SW-UNet was 4.9 times smaller than 2D-FFT based wrinkle segmentation approach and was 2.9 times smaller than U-Net.
As a demonstration, we compared the contractile force of U2OS cells and showed that cells with mutant KRAS gene exhibit larger force compared to the wild-type cells.
Our new machine learning based algorithm provides us an efficient, automated and accurate method to compare the cell contractile force.
We believe that our network SW-UNet and CNN building strategy would be useful for other applications.

\section*{Acknowledgement}
This work was supported by JSPS KAKENHI Grant Number 18H03518. 

\bibliography{reference}

\end{document}